# Label-free bacteria detection using evanescent mode of a suspended core terahertz fiber

Anna Mazhorova,[1] Andrey Markov,[1] Andy Ng,[2] Raja Chinnappan,[2] Mohammed Zourob[2,*] and Maksim Skorobogatiy[1,*]

[1]*École Polytechnique de Montréal, Génie Physique, Québec, Canada*
[2] *Institut National de la Recherche Scientifique, Varennes, Québec, Canada*
[*]*maksim.skorobogatiy@polymtl.ca;*

**Abstract:** We propose for the first time an *E. coli* bacteria sensor based on the evanescent field of the fundamental mode of a suspended-core terahertz fiber. The sensor is capable of *E. coli* detection at concentrations in the range of $10^4$-$10^9$ cfu/ml. The polyethylene fiber features a 150 μm core suspended by three deeply sub-wavelength bridges in the center of a 5.1 mm-diameter cladding tube. The fiber core is biofunctionalized with T4 bacteriophages which bind and eventually destroy (lyse) their bacterial target. Using environmental SEM we demonstrate that *E. coli* is first captured by the phages on the fiber surface. After 25 minutes, most of the bacteria is infected by phages and then destroyed with ~1μm-size fragments remaining bound to the fiber surface. The bacteria-binding and subsequent lysis unambiguously correlate with a strong increase of the fiber absorption. This signal allows the detection and quantification of bacteria concentration. Presented bacteria detection method is label-free and it does not rely on the presence of any bacterial "fingerprint" features in the THz spectrum.

**1. Introduction**

The selectivity and sensitivity of optical biosensors have put them into the class of most popular biosensors [1-4]. A large number of methods for highly sensitive chemical or biological detection that operate in the THz range were developed. Among these methods, metamaterials and frequency selective surfaces with a resonant frequency response that is tunable by design have been employed. It has been shown that even a small quantity of material deposited on THz metamaterial can shift its resonance frequency [5]. In addition, techniques based on THz vibrational spectra characterization have become available for biological materials in the form of the diluted water solutions and pressed pellets [6,7]. Recently, molecular spectroscopy of lactose dispersed on the top of metal wire was demonstrated [8]. A related technique [9] measured the absorption properties of liquids in the THz wave range, based on the interaction between the liquid and the evanescent wave covering the surface of a cylindrical waveguide made of high-resistivity silicon. A similar label-free THz sensing based on the plastic wires was also reported [10].

The present work is motivated by the recent reports on evanescent field or surface plasmon resonance (SPR) biosensors in the visible and near-IR regions [10,12]. Most of the investigations in these spectral regions focus on the detection of relatively low concentrations of proteins and small molecules with sizes about/less than tens of nanometers. The reason is that the probing length of an evanescent field of an optical waveguide, or the size of the tail of a plasmon that are used to detect changes in the optical properties at the interfaces, is deeply subwavelength. For example, a typical size of a plasmon tail in the visible range is smaller than 100 nm, which allows us to track the dynamics of a biochemical event in the immediate vicinity of the interface. However, when one tries to measure directly the presence of a bacteria on a metallic surface, the standard SPR method runs into a difficulty as the size of the probing field (plasmon) is not only smaller than the size of a bacteria, but is also comparable to the size of a phage molecule (~100 nm) that connects the bacteria to the surface. Therefore the detection of large target such as cells and bacteria by SPR methods remain a challenge. This results in the probing of a phage layer rather than a bacteria layer. Therefore, it is desirable to match the size of the measured object to that of a probing field. In the case of bacteria detection with a typical bacteria size of 1-10 μm, one can pursue two approaches. One is to use short wavelengths (such as in the visible or near-IR) [13], and then use well delocalized evanescent field of a waveguide mode (for example, mode of a subwavelength fiber) to cover the object. Another alternative is to use longer wavelengths (such as in the mid- or far-IR) and then match the extent of the modal evanescent field by the waveguide design.

In this paper we explore the possibility of using THz fibers operating at wavelength longer than the size of the studied object. We use the high refractive index combination of core (polyethylene) and cladding (air) in order to significantly reduce the size of the probing evanescent tail. This allows us to develop a label-free technique (target bacteria are not labeled or altered and are detected in their natural forms) that is based on the interaction between the evanescent wave localized around the core of the fiber and a thin layer of bacteria covering it. This technique is then used for the detection and quantification of *E. coli* bacteria in the THz spectral range.

In addition, we established a measurement protocol that allows measurement of biological properties in the aqueous environment, while using THz waves that are normally unsuitable for this purpose because of the strong absorption of water. This so-called "frozen dynamics" approach allows all the biological processes to take place in water; while the system is robust enough to allow drying for THz measurements at the intermediate stage without affecting the biological subcomponents (such as phages).

## 2. THz subwavelength fiber

The major complexity in the design of the terahertz waveguides is the fact that most dielectric materials are highly absorbing in this spectral region [14] with a typical loss in excess of ~1 cm$^{-1}$. To combat the material losses, a novel approach based on the introduction of porosity in the fiber core has been recently introduced by our group [15]. We have proposed that since the absorption loss is lowest in dry air, one way to reduce waveguide propagation loss is to maximize the fraction of power guided in the air.

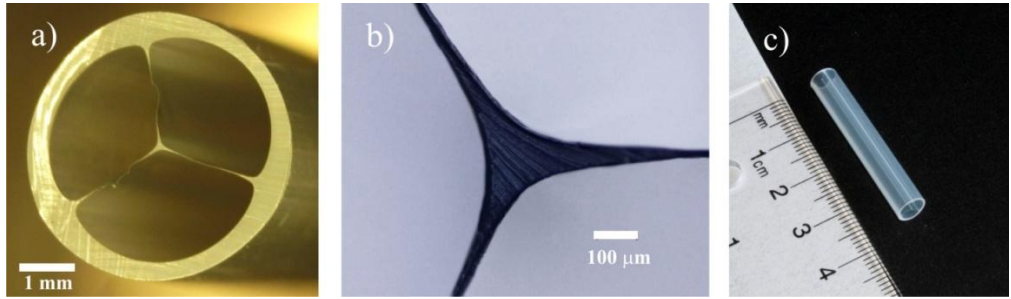

Fig. 1. (a, b) The THz fiber featuring a 150 μm core suspended by three 20 μm-thick bridges in the center of a 5.1 mm diameter tube, (c) 4 cm-long fiber piece used in the experiments.

One of the waveguides that operates on this principle was recently described in [16] and presents a subwavelength fiber suspended on thin bridges in the middle of a larger protective tube (see Fig. 1). Large channels formed by the bridges and a tube make a convenient opto-microfluidic system that is easy fill with liquid analytes or purge with dry gases. Particularly, the THz subwavelength waveguide used in our experiments features a 150 μm core fiber suspended by three 20 µm-thick bridges in the center of a 5.1 mm diameter tube (Fig. 1. (a, b)) of 4 cm in length. This waveguide design presents several important advantages for bio-sensing applications. First, the waveguide structure allows direct and convenient access to the fiber core and to the evanescent wave guided around it. Second, the outer cladding effectively isolates the core-guided mode from the surrounding environment, (e.g., fiber holders), thereby preventing the undesirable external perturbations of the terahertz signal.

All the data in our experiments was acquired using a modified THz-TDS (Time-Domain Spectroscopy) setup. The setup consists of a frequency-doubled femtosecond fiber laser (MenloSystems C-fiber laser) used as a pump source and identical GaAs dipole antennae used as source and detector yielding a spectrum ranging from ~0.1 to 3 THz. Contrary to the standard THz-TDS setup where the configuration of parabolic mirrors is static, our setup has mirrors mounted on translation rails. This flexible geometry facilitates mirrors placement, allowing measurement of waveguides up to 45 cm in length without realigning the setup. Fig. 2(a) illustrates the experimental setup where the fiber is fixed between the two irises and placed between the focal points of the two parabolic mirrors.

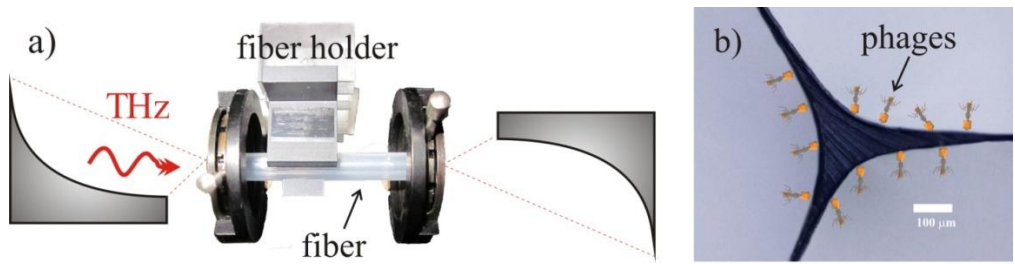

Fig. 2. (a) Schematic of the experimental setup: fiber is placed between the focal points of the two parabolic mirrors, (b) schematic presentation of phages adsorbed onto the fiber core. The capsid adsorbed onto the fiber, while the tail (which is specific to the bacteria) faces towards the cladding for bacteria capturing.

## 3. Materials and methods

*3.1 Bacteria culture*

Bacteria solution was prepared as following. A frozen stock of *E. coli* B strain was used to seed an overnight culture in LB media (liquid media for growing bacteria, contains 10g Tryptone, 5g yeast extract, 10 NaCl in 1L $H_2O$). Bacteria were harvested by centrifugation at 3000xg for 10 min, followed by washing in PBS (Phosphate Buffer Saline). Dilutions of the overnight culture were spread on LB agar plate to determine the titre, expressed in colony forming unit (cfu). Appropriate dilutions of bacteria stock with a concentration $10^4$ – $10^9$ cfu/ml were made in PBS for subsequent experiments.

*3.2 Phage production*

100uL of *E.coli* B log-phase culture was added to 3ml cooled top agar and poured onto LB agar plate until solidified. 100uL of phage stock specific to *E.coli* B was added onto the solidified top agar and incubated at 37ºC overnight. A macroplaque was developed on the lawn of bacteria after incubation. 3ml of lambda buffer was added onto the top agar, which is then scraped off and collected in a 50ml centrifuge tube, followed by an additional washing of the LB agar plate with 3ml lambda buffer. 3 drops of chloroform were added into the suspension, vortexed and centrifuge at 3500xg for 10min to pellet pieces of agar and bacteria. The supernatant was then passed through a 0.22um filter to remove the remaining bacteria. The filtrate is the stock phage solution and is prepared fresh for each experiment. The titre of the phage stock was determined by serial dilution of the stock, followed by infection to *E.coli* as described above. Plaques were counted and the titre of the phage stock is expressed in plaque forming unit (pfu).

## 4. Characterization

The light propagating through an optical fiber consists of two components: the field guided in the core and the exponentially decaying evanescent field in the cladding. For evanescent sensing one typically removes the fiber cladding and introduces analyte into the immediate proximity of the fiber core [17]. In the presence of strong absorption or scattering in the analyte, the intensity of guided wave would decrease, which can then be then used to interpret various analyte properties. In application to THz evanescent sensing, both analyte refractive index and analyte absorption can be deduced by comparing the signals from the empty waveguide with that from the waveguide containing a layer of the measured material. By measuring the changes in both amplitude and phase caused by the addition of the analyte layer even very small amounts of analyte can be detected and characterized (see, for example, [18] where nanometer-size aqueous layers were characterized in THz).

An immediate complication for evanescent biosensing in THz is that natural environment for a majority of bacteria is water and water is highly absorbing in THz spectral range. To avoid this problem, we have developed a protocol where all the bio-reactions take place in the natural aqueous environment, while the THz measurement is done on dried samples.

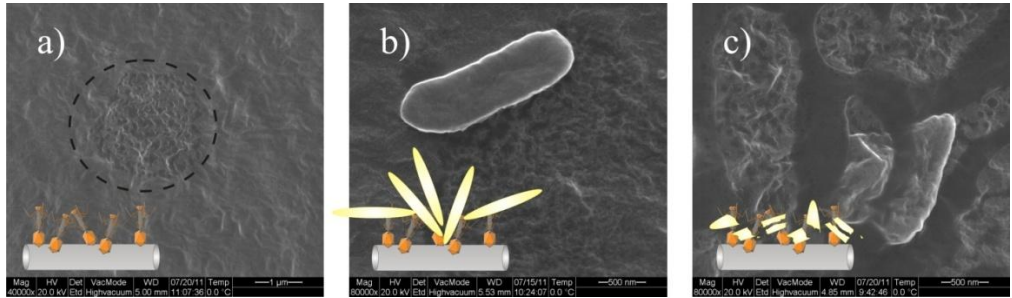

Fig. 3. SEM images illustrating each step of the experiment (a) step 1- phages are immobilized on the fiber core surface, (b) step 2 – capturing of *E.coli* bacteria by the phages, and lysis of the bacteria (c) step 3 - fiber is washed with PBS, bacteria chunks remain bound to the core surface.

In what follows all the SEM images were taken with a Quanta 200 FEG scanning electron microscope operating in high vacuum mode. First, samples were fixed with formalin for 12 hours and then metalized with several nm-thick layer of gold.

To follow the dynamics of the bacteria binding we performed several experiments according to the three-step procedure described below. Before the measurement, the suspended core fiber (4 cm in length) was cleaned with isopropyl alcohol and distilled water. The fiber was incubated overnight (12 hours) with T4-bacteriophage solution at $10^{10}$ pfu/ml concentration. During this incubation, phages are adsorbed non-specifically onto the plastic fiber surface. The T4 phages recognize and bind to specific cell surface protein on the *E.coli* bacteria using their tail proteins (scheme of the process is shown at Fig. 2 (b)). This recognition is highly specific and has been extensively used for typing of *E. coli*.

After phage adsoprtion the fiber was washed 3 times with distilled water and PBS. Only 10% of the fiber's bridges surface and core were covered by phages. The exposed fiber surface was therefore blocked with bovine serum albumin (BSA) for 30 min in order to reduce non-specific adsorption of the bacteria, and washed again with the buffer solution and dried out. Blocking with BSA is required because it is not possible to achieve 100% coverage of the fiber by the phages. Without BSA blocking, exposed fiber surface could cause non-specific adsorption of the bacteria and other components present in the sample, causing false positives.

SEM imaging was performed to confirm the adsorption of the phages onto the fiber core as well as the stability of the phage layer, and the extent of fiber surface coverage. Fig. 3 (a) illustrates phages adsorbed onto the fiber (dark spots within the area define by dashed line). At the same time, THz transmission through the fiber was measured. Fiber transmission spectrum is shown in Fig. 4 (a), black line.

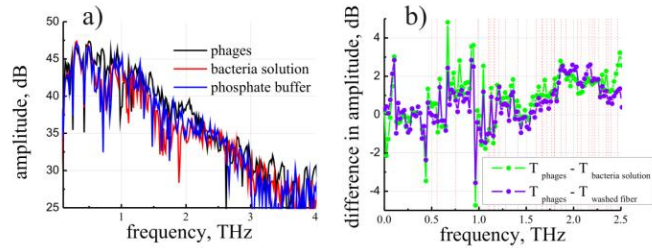

Fig. 4. (a) Transmission spectrum of the fiber during each step of the experiment: step 1- black line, only phages, step 2 – red line, transmission of the fiber decreased due binding of *E. coli* bacteria to the phages, step 3 – blue line, fiber is washed with PBS. Fiber transmission increased but not up to the level of the first step, suggesting that some bacteria (or parts of it) remain bound to the fiber via specific interaction with the phages.(b) difference in transmission between each step of the experiment.

After addition of the bacteria and subsequent binding of the phage on the bacterial surface, the phage infection process will start. Phage progeny will accumulate and eventually the bacteria host is ruptured followed by the release of the phage. The period between infection and bacteria lysis is known as the latent period and usually last for 25 minutes. The second measurement of the THz transmission through the fiber was done during the first 15 min after the introduction of bacteria (Fig. 4 (a), red line). It could be clearly seen that the transmission of the fiber decreases considerably due to binding of the bacteria to the phages. A zoom-in of the waveguide core after some reaction time ~5-10min is shown in Fig. 5. There, the dashed line circumscribes the part of a core covered with the bacteriophages; *E. coli* Bacteria are clearly visible bound to the region covered by the phages and not to the rest of the core surface blocked with BSA.

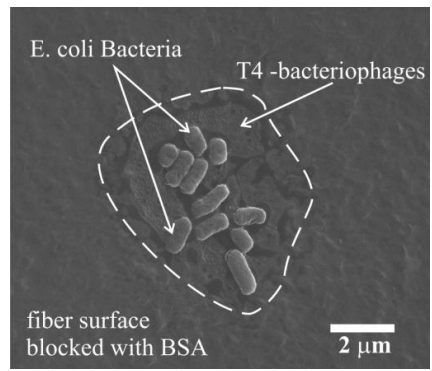

Fig. 5. Zoom-in of the waveguide core, dashed line marks the area covered with bacteriophages. Bacteria are clustered in the phage-covered surface with the rest of the surface blocked by BSA.

In the last step, after 25 min of reaction time, the fiber was washed with the PBS in order to remove the unbound bacteria on the fiber surface and then again dried out. Fiber transmission (Fig. 4 (a), blue line) increased compared to the previous step, but not up to the level of the first step, suggesting that some bacteria (or parts of them created by lysis and rapture) remain bound to the fiber via specific interaction with the phage.

The following SEM images (Fig. 6) were prepared at different moments of time during the process of specific recognition and binding of the *E. coli* bacteria with phages in order to understand the process that take place during the last 10 min of the 2$^{nd}$ step of the measurements.

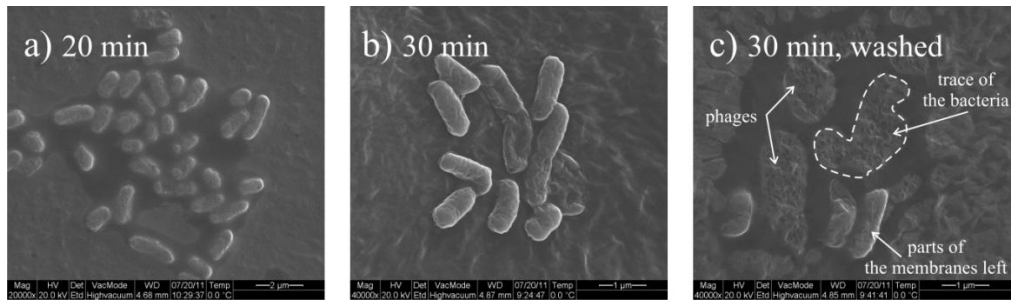

Fig. 6. SEM images of the fiber core (a) after 20 min since the beginning of the 2nd step; (b) after 30 min before washing with PBS (end of the 2nd step); (c) after 30 min after washing with PBS. The bacteria shape changed from a uniform rod shape to a random shape. Eventually, the bacteria cell wall ruptures and releases intracellular components with only the cell membrane left on the fiber.

Fig. 6 (a) shows fiber core after 20 min of interaction of *E. coli* bacteria with the phages. The bacteria are bound to the phages and have well defined rod shapes. Fig. 6 (b) shows the fiber core at the end of the 2$^{nd}$ step, after 30 min of bacteria interaction with pages. It is clearly seen that by the end of the 2$^{nd}$ experiment the bacteria change its shape form the rod-like to a highly distorted rod shape which is a result of bacteria lysis. During lysis the bacteria cell wall eventually ruptured and releases intracellular components with only the cell membrane left on the fiber. Fig. 6 (c) illustrates the core of the fiber after 30 min of interaction of the bacteria with the phages and after washing with PBS. After the fiber was washed, traces of the bacteria (membrane fragments) can be clearly seen still bound to the phages. Visually, washed fibers corresponding to Fig. 6 (c) look cleaner than the same fibers before washing.

In order to explain a considerable increase in the fiber transmission loss due to bacteria binding (during the course of the 2$^{nd}$ measurement) we note that *E. coli* consists largely of water. Therefore, it is expected that its mobilization on the core surface would leads to a significant loss increase due to water absorption. Particularly, cell dry weight ("dry weight" is the remaining material after all the water is removed) of the *E. coli* bacteria is only 30% and can be subdivided into protein (55%), RNA (22.5%), DNA (3.1), lipids (9.1%), peptidoglycan (2.5%), polysaccharides (3.4%) and Glycogen (2.5%), while remaining 70% of the bacteria weight is water [20]. Additionally, we expect that presence of the large and irregular bacteria clusters on the fiber surface would lead to scattering loss.

Similarly, after washing the fiber at the end of the 2$^{nd}$ measurement, fiber transmission is expected to increase as only the dry chunks of bacteria membrane remain on the surface. Moreover, their surface coverage is smaller than this of the bacteria before washing, therefore scattering loss is expected to be smaller in washed fibers. For completeness, we note that the outer membrane of the gram-negative bacteria such as *E. coli* is a complex structure composed of phospholipids, proteins, and polysaccharides that controls the permeability and helps maintain the shape and rigidity of the cell. The protein complement of the outer membrane includes a Murein-lipoprotein that is one of the most abundant proteins in bacteria: there are about $7.2 \cdot 10^5$ molecules per cell.

Finally, from Fig. 4 we observe that there are two spectral regions in the vicinity of 0.7 THz and in the range of 1.5-2.0 THz where changes in the fiber absorption are especially pronounced. These particular spectral regions of high sensitivity were consistently identified in each of the several repetitions of our experiments. The physical nature of high absorption in these spectral regions is still unclear to us, while there is an indication that it can be related to spectral signatures of specific macromolecules abundant in the *E. coli* bacteria composition. In particular, in the THz spectral range large tandem repeats peptides or nucleosides have some observable spectral features, whereas nonperiodic biological molecules such as DNA and proteins have relatively featureless spectra [2]. Optimal regions for the THz detection of

*E. coli* bacteria found in our work correlate with the position of the absorption peaks of the Murein-lipoprotein or Broun's lipoprotein (BLP), that is one of the most abundant proteins in bacteria [21,22]. Additionally, there is a 0.68 THz absorption peak of Thioredoxin protein which a commonly occurring electron transport protein in a bacteria. For the sake of clarity we also note that all the measurements were done in the nitrogen purged cage at 12% of humidity, therefore all the narrow dips in the transmission spectrum as observed in Fig. 4 (for example the 0.752 THz peak) are caused by the water vapor [23,24] and has to be distinguished from the broad absorption features due to bacteria.

## 4. Results and Discussion

In this section we present and interpret changes in the fiber absorption loss during bacteria binding process. Fiber absorption loss is defined as $\alpha = -\ln(P/P_{ref})/L$ where $L$ is the fiber length, $P_{ref}$ is the transmitted power through the setup without fiber, and $P$ is a transmitted power through a setup with fiber. To see the influence of bacteria binding on the THz transmission of a suspended core fiber in dynamics we measure absorption losses of the fiber (with a dry sample) as a function of time. Particularly, during each step of the experiment we did at least 10-15 scans of the transmitted spectrum (one scan takes ~1 min), as a result, we have absorption losses of the fiber as function of time. Also, as it was mentioned before, the THz measurement is done with dried samples, while the bio-reactions take place in the natural aqueous environment. In particular, after 1st step of the experiment fiber was incubated with bacteria aqueous solution and then after 15 min (during this time phages have already captured bacteria) fiber was dried out for the transmission measurements. This gives us ~15 min gap in the sensogram where it was impossible to measure losses due to high water absorption in THz spectral range.

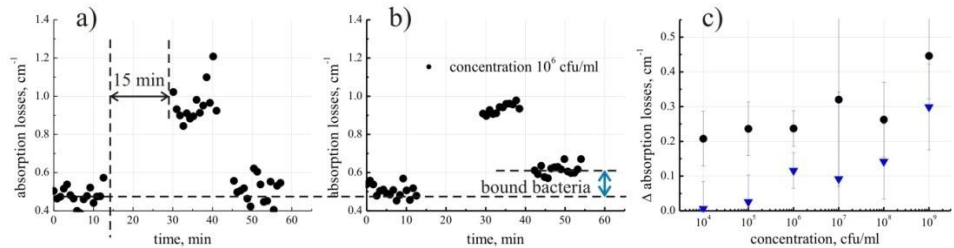

Fig. 7. Absorption losses of the fiber. (a) reference sensorgram; (b) sensorgram for bacteria concentration at $10^6$ cfu/ml as a function of time; (c) Correlation between the changes in the fiber absorption losses and bacteria concentration: difference between base level (absorption losses of the fiber with phages immobilized on the fiber core) and losses of the "washed" fiber is shown in blue triangles, as a function of a concentration. Black dots correspond to the difference between absorption losses of the fiber after 2nd and 3rd steps (before and after fiber washing).

Figure 7 illustrates effect of the bacteria binding on the THz transmission of a suspended core fiber at 0.7 THz. First the reference sensorgram was acquired (Fig. 7(a)) for which the fiber was prepared exactly the same way as described in the characterization section, however, no phages were used (phage solution was simply substituted with buffer during the 1st step), while the fiber surface was blocked with BSA to prevent non-specific bacteria binding to the core surface. Absorption losses returned to the initial level after bacteria were washed away with PBS. This transmission measurement and SEM images (not showed here), prove that the fiber surface was indeed blocked with Bovine Serum Albumin and no non-specific adsorption of the bacteria took place on the fiber surface.

In Fig. 7(b), results of the experiments with phages are presented where despite extensive washing, some bacteria are always retained by specific binding to the phage-coated fiber, and as a consequence, fiber absorption loss does not return to the original level. More importantly, we are able to show the correlation between the fiber absorption losses and the bacteria concentration of the samples, further suggesting the specificity of the detection method. Thus, in Fig. 7 (c) difference between the base level (absorption losses of the fiber with phages immobilized on the fiber core) and losses of the "washed" fibers are shown in blue triangles, as a function of bacteria concentration in the solution used during the $2^{nd}$ experiment. Also, in black circles we show difference between absorption losses of the fiber after $2^{nd}$ and $3^{rd}$ steps (before and after washing of the fiber). This difference is related to a signal due to both specific bacteria binding to the phages and nonspecific temporary adsorbed bacteria onto a fiber surface so by itself it cannot be used for sensing. From the presented data (blue triangles) we see that the detection of limit of this method is around $10^4$ cfu/ml, which is comparable with existing commercially available methods [13] The commercial methods usually employ antibodies (Abs) as recognition elements instead of bacteriophages. Production of the specific Abs, however, is difficult, expensive and very time-consuming. Also, antibodies easily lose their activity when subjected to changing environmental conditions. Bacteriophages are alternative method of bacteria capture that offer many advantages compared to antibodies. Particularly, phages are easy to produce in large quantities at a relatively low cost, they have long storage life and unlike antibodies can be physically adsorbed onto the fiber surface.

Considerable sensitivity enhancement of our method is possible by increasing the amount of the initial phage coverage of the fiber surface by better anchoring the phages via covalent immobilization onto the fiber surface, rather than physical adsorption. This however implies considerable effort in fiber surface biofunctionalisation chemistry which is beyond the scope of this paper.

## 5. Simple theoretical model to explain changes in the fiber absorption loss

We believe that strong changes in the fiber transmission loss during the $2^{nd}$ phase of the experiment (before washing the fiber) can be explained by the presence of water containing bacteria on the fiber core due to non-permanent adsorption or sedimentation during rapid drying of samples.

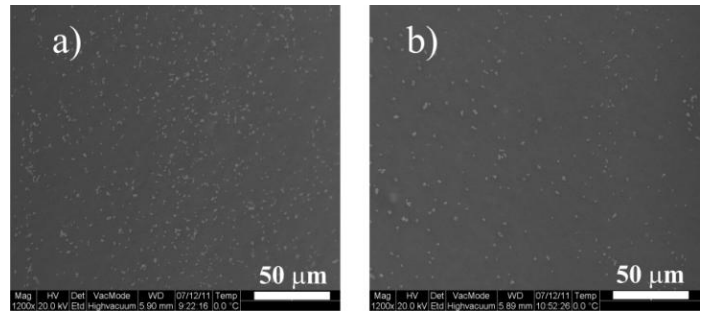

Fig. 8. SEM images of the fiber core with bacteria concentration of (a) $10^9$ cfu/ml and 10 min interaction time (b) with a concentration of $10^6$ cfu/ml, 10 min interaction time.

As seen in Fig. 7(c) these losses are proportional to the bacteria concentration. Moreover, from the direct observation of the SEM images, bacteria coverage of the fiber core is also proportional to the bacteria concentration. For example, at $10^9$ and $10^6$ cfu/ml, the average coverage was correspondingly 4.35% and 1.52% (Fig. 8. (a,b). These values were achieved after analyzing presented SEM images via custom LabVIEW software. First we determined the ratio of the area covered with the bacteria (in pixels) to the total area of the picture in percentage. At the same time total number of the bacteria in the field of view was also counted, which divided by the area of the SEM image yields the coverage ratio of 45 cells/100 $\mu m^2$ for the concentration $10^9$ cfu/ml and 15 cells/100 $\mu m^2$ for the concentration $10^6$ cfu/ml. Despite the fact that at concentration $10^9$ cfu/ml we have 1000 times more bacteria then at the concentration $10^6$ cfu/ml, we don't have a significant increasing of the bacteria coverage ratio due to limitation of the initial phage coverage of the fiber surface (only 10%) and number of phages required for the bacteria binding and subsequent lysis. As we will see in the following, it is the presence of the water containing bacteria on the fiber surface that leads to strong increase in the fiber absorption before washing. After washing, only dry bacteria chunks remain bound at the fiber surface and, therefore, fiber loss decreases. Thus, the additional loss after washing is most probably due to scattering on the bacteria remains.

To create theoretical model of the experiment we first studied the modal structure of our fiber. We started by importing the fiber cross-section as captured by the optical microscope Fig. 1. (a) into COMSOL Multiphysics Finite Element software. Complex effective refractive indices and field profiles of both core guided and cladding modes were then studied. Then, the power coupling coefficients have been computed from the overlap integrals of the modal fields with those of the Gaussian beam of a source as described in [16]. Inspection of the coupling coefficients confirms that the fundamental mode is predominantly excited in the 0.2 – 2.0 THz frequency range. Transverse distribution of the longitudinal power flux in the fundamental mode of a fiber at 0.71THz is shown in Fig. 9 (a) and a typical penetration depth of the evanescent field into the air cladding is determined to be $25 \pm 5 \mu m$ which is in the good agreement with a theoretical estimate $\lambda / \left(4\pi \sqrt{\varepsilon_{core} - \varepsilon_{cladding}}\right)_{0.71THz} \approx 29 \mu m$.

Next, absorption losses of a suspended core fiber with a bacteria layer have been investigated numerically using COMSOL. Simplified fiber geometry has been used to facilitate the task of distribution of bacteria on the core and characterization of the bacteria coverage ratio. We have reduced the transverse shape of the core to a circle and we have assumed that the bridges are straight and of constant width. The values of the waveguide core radius and the bridges width were chosen to make the changes in the effective refractive index, the absorption losses and the modal size minimal compared to the microscope imported design, and, hence, for simulation we used $D_{core} = 175 \mu m$ and $d_{bridges} = 30 \mu m$.

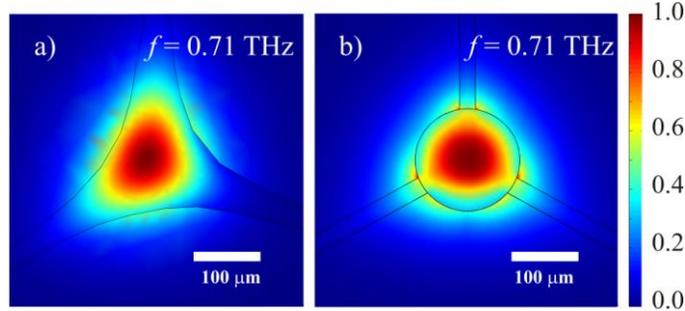

Fig. 9. Transverse distribution of power flow for the fundamental mode of (a) the real waveguide and of (b) the simplified design of the waveguide profile. The field is confined in the central solid core and is guided by total internal reflection.

*E. coli* are typically rod-shaped, about $2.0 \mu m$ long and $0.5 \mu m$ in diameter. To simulate bacteria we have approximated their shape as an infinite cylinder of diameter $d_{bacteria} = 1 \mu m$ positioned along the fiber length. Considering that bacteria are mostly made of water (70% by total wet wt. [20]) we considered their refractive index and losses as those of the bulk water. The complex dielectric permittivity of water has been calculated using the full Rocard-Powles-Lorentz model with the parameters taken from [26]. For 0.71 THz the water refractive index is 2.19+0.65i resulting in absorption loss of 191 cm$^{-1}$. For low density polyethylene the real part of the refractive is about 1.514 in THz frequency range and the material loss are computed from the measurements [25] $\alpha \left[cm^{-1}\right] = 0.75 \cdot v^2 - 0.1 \cdot v + 0.2$ where $v$ is the frequency in THz.

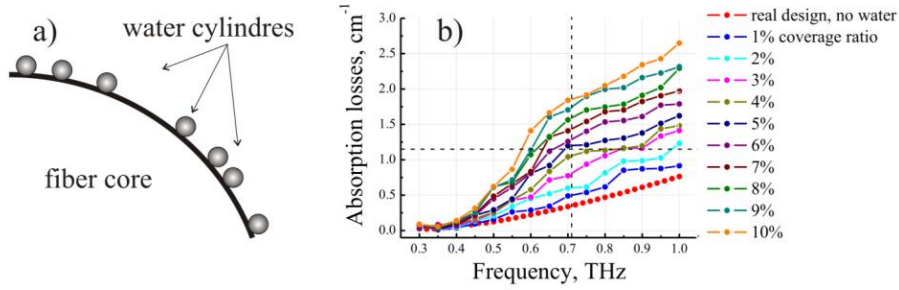

Fig. 90. (a) Randomly distributed 1 μm water cylinders on the waveguide core's surface for a given value of surface coverage. (b) Absorption loss of the fundamental mode for the fiber with a bacteria layer as a function of the bacteria coverage ratio. The crossing of the dash lines in the figure corresponds to the experimentally measured value of the propagation losses minus the theoretically estimated coupling loss for the $10^9$ cfu/ml concentration of bacteria.

From the SEM images of the fiber surface with bacteria (see Fig. 8) we see that bacteria are randomly positioned on the core surface with relatively low concentrations. To take this into account in our simulations the water cylinders (bacteria) were randomly distributed (see Fig. 10 (a)) on the fiber core surface with a fixed value of a coverage ratio (i.e., the core's surface occupied by the water cylinders to the overall core's surface). Then, the fundamental mode and its losses were found using FEM software, and the modal losses were then averaged over 10 random bacteria distributions. In Fig. 10 (b) we present thus calculated propagation losses of the fiber fundamental mode as a function of the bacteria coverage ratio. We observe that increase in the bacteria coverage ratio leads to a linear growth of the modal absorption losses. Due to much higher absorption losses of water compared to those of a polyethylene core, even a small coverage ratio of the bacteria results in large increase in the modal absorption loss. As an example, in Fig. 10 (b) we show at the intersection of the dashed lines the case of bacteria of concentration $10^9$ cfu/ml that corresponds to the experimental fiber loss of ~1.2cm$^{-1}$ at 0.71THz. Our theoretical analysis predicts that this loss is produced by the absorption loss of the fundamental modes due to 4% coverage ratio of water on the core's surface multiplied by the modal coupling coefficient from the Gaussian beam to the fundamental mode, which corresponds well with the experimental value of the bacteria coverage ratio 4.35% found from the SEM observations

Another important feature found in Fig. 10 (b) is an observation that for a given water coverage ratio, fiber absorption loss is small at low frequencies, then shows a rapid increase in the vicinity of 0.5THz, and, then, at frequencies above 0.7THz it simply follows the absorption loss of the polyethylene material with an almost constant frequency independent loss contribution from the water layer. This is easy to rationalize by noting that at low frequencies (<0.3THz in our case) mode of a subwavelength fiber is strongly delocalized, and, therefore, its presence in the thin water layer around the core is minimal. When frequency increases (~0.5THz in our case), fiber mode shows rapid localization in the vicinity of a core region, and, therefore, its relative presence in the water layer and, hence, losses also rapidly increase. This also explains why in Fig. 4, there is little difference in the measured transmission losses of bare and bacteria activated fibers at frequencies below 0.5THz. Finally, at even higher frequencies (above 0.7 THz in our case) fundamental mode becomes mostly localized in the fiber core. Absorption loss contribution due to a small water layer on the fiber surface decreases and becomes almost independent of the operation wavelength because the decreasing ratio of power inside water layer is compensated by the increasing with frequency water absorption losses. Finally, at even higher frequencies (above 0.7 THz in our case) fundamental mode becomes mostly localized in the fiber core of radius $R$ with a fraction of power in thin layer (of thickness $d$) on the core surface being $\sim \lambda^2 d / R^3$. Here we assume that the layer thickness is smaller than the evanescent field penetration depth into the air cladding

$d < \lambda / \left(4\pi\sqrt{\varepsilon_{core} - \varepsilon_{clad}}\right)$, which is a true assumption all the way up to 10 THz. Considering that the bulk absorption loss of water increases $\sim \nu$ with frequency [25] we conclude that at higher frequencies absorption loss contribution due to a small water layer on the fiber surface becomes a slowly decreasing function of the operation frequency $\sim d/(\nu R^3)$, which should be compared to the $\sim \nu^2$ behavior of the fiber loss due to polyethylene absorption (see Fig. 10).

## 6. Conclusion

We demonstrate the possibility of using suspended core (*d=150 μm*) polyethylene THz fibers for sensing of the *E. coli* bacteria with detection limit of $10^4$ cfu/ml. Structure of the suspended core fiber allows convenient access to the fiber core and to the evanescent part of the guided wave. Outer cladding effectively isolates the core-guided mode from interacting with the surrounding environment such as fiber holders and humid air, thus making the fiber convenient to handle and to operate.

It was shown that selective binding of the *E. coli* bacteria to fiber surface bio-functionalized with specific phages unambiguously influences the THz transmission properties of the suspended core fiber. Moreover, changes in the fiber absorption loss can be correlated with the concentration of bacteria samples. Thus, our setup allows not only detection of *E. coli*, but also quantitative measurement of its concentration. Presented bacteria detection method is label-free and it does not rely on the presence of any bacterial "fingerprint" features in the THz spectrum.

Simple theoretical model was also developed in order to explain observed changes in the fiber absorption losses. It was found that strong increase in the fiber loss can be explained by absorption of a thin water layer corresponding to the bound bacteria.